\documentstyle[preprint,aps]{revtex}

\begin{document}

\draft

\title{Updated analysis of $\pi$N elastic scattering data to 2.1~GeV: \\
       The Baryon Spectrum}
\author{Richard A. Arndt, Igor I. Strakovsky$^\dagger$ and Ron L. Workman}
\address{Department of Physics, Virginia Polytechnic Institute and State
University, Blacksburg, VA 24061}
\author{Marcello M. Pavan}
\address{University of British Columbia, Vancouver, British Columbia,
Canada V6T 2A3}

\date{\today}
\maketitle

\begin{abstract}
We present the results of energy-dependent and single-energy
partial-wave analyses
of $\pi$N elastic scattering data with laboratory kinetic energies below
2.1~GeV.  Resonance structures have been
extracted using Breit-Wigner fits, speed plots,
and a complex plane mapping of the associated poles and zeroes.
This is the first set of resonance parameters from a VPI analysis
constrained by \hbox{fixed-t} dispersion relations. We have searched our
solutions for structures which may have been missed in our previous
analyses, finding candidates in the $S_{11}$ and $F_{15}$ partial-wave
amplitudes. Our results are compared with those
found by the Karlsruhe, Carnegie-Mellon$-$Berkeley, and Kent State groups.

\end{abstract}
\vspace*{0.5in}

\narrowtext
\section{Introduction}
\label{sec:intro}

We have performed a partial-wave analysis of pion-nucleon elastic scattering
data up to a laboratory pion
kinetic energy of 2.1~GeV. This work supersedes our last published
analysis~\cite{ar91} (named SM90). The present
analysis (called SM95) was performed on a larger data base,
and was constrained by
fixed-t dispersion relations (FTDR).
In a previous study~\cite{ar94} (solution FA93) employing
FTDR, we focused on a determination
of the pion-nucleon coupling constant ($g^2/4\pi$), finding the value
$g^2/4\pi = 13.75 \pm 0.15$. In the present study we
concentrate on the baryon spectrum as determined by Breit-Wigner fits,
speed plots, and complex plane mappings.
As our algorithm for implementing  FTDR constraints has been described
in Ref.\cite{ar94}, we will only outline the method in this paper.
One further change in our method of analysis was made in response to
a suggestion made by H\"ohler\cite{privcomm}.
We have scanned our energy-dependent
solution for ``missing" structures by sweeping an
adjustable Breit-Wigner resonance contribution through each partial-wave.
As a result, we have found some evidence
for a small number of additional structures.

In Section II, we will  briefly describe the additions made to our
database since the publication of Ref.~\cite{ar91}.
In Section III, we will review the basic formalism\cite{ar91,ar94,ar85}
used in our analyses.
Results for the baryon spectrum and associated couplings will be given in
Section IV. Here we will also compare the present solution with the
older solution SM90.
Finally, in Section V, we will compare our resonance spectrum with
the results of the Karlsruhe\cite{ho83,ho92,ho94},
Carnegie-Mellon$-$Berkeley (CMB)\cite{ke79},
and Kent State\cite{ma92}
groups. In particular, we will
comment on discrepancies in the observed resonance states.

\section{The Database}
\label{sec:dat}

Our previous published $\pi N$ scattering analysis\cite{ar91} (SM90)
was based on 10031
$\pi ^{+}p$, 9344 $\pi ^{-}p$, and 2132 charge-exchange data.
Since then we have added 358 $\pi ^{+}p$, 710
$\pi ^{-}p$, and 53 charge-exchange data. Some other measurements were
removed\cite{flag} from the analyses in order to resolve database conflicts.
The new low-energy $\pi N$ data were produced mainly at the
TRIUMF, LAMPF, and PSI meson facilities, and at the SPNPI and ITEP
facilities in the 1~GeV region.
The distribution of recent (post-1990) data is given schematically in Fig.~1.

Since most of the new data~\cite{ab91p}-\cite{wi93p} are from high-intensity
facilities, they generally have smaller statistical errors and thus have
greater influence on the fits. A large fraction of the new
$\pi ^{\pm}p$ data were produced at energies spanning the $\Delta$ resonance.
TRIUMF has produced differential cross sections with an
accuracy of 1--2\% \cite{br94p} and partial total cross
sections\cite{fr91,fr93}.  LAMPF has produced
a set of polarization parameters $P$, $R$, and
$A$ \cite{su93}.
TRIUMF and LAMPF have produced total\cite{fr93} and differential cross
sections\cite{is93p,po93p}, and analyzing
powers\cite{st93p} for the charge exchange reaction.
After a revised analysis and energy calibration,
the Karlsruhe group, working at PSI, has provided
a final set of both forward\cite{me93p}
and backward differential cross sections\cite{jo93p} and analyzing
powers\cite{wi93p} at low energies.

Most of new $\pi ^{\pm}p$ differential cross sections and analyzing powers
above 780~MeV were measured at ITEP\cite{ab91p,al91}.  Some proton
spin rotation parameters were measured below 600~MeV at
SPNPI\cite{lo94} and at 1300~MeV at ITEP\cite{su94p}.

Other experimental efforts will soon provide data in the low to
intermediate energy region. A precise measurement of
${\pi}^{\pm}p$ elastic scattering cross sections was made in
experiment ({\bf E645}). This experiment covered
the $\Delta$ isobar region and was completed
at TRIUMF in the Summer of 1992\cite{pa93p}. Partial total cross
section measurements  ({\bf E1190}) for angles greater than 30$^{\circ}$
(lab) have been made at LAMPF in Summers of 1991 and 1992\cite{kr93p}.
Data was taken between 40 and 500~MeV for $\pi ^{+} p$
and between 80 and 300~MeV for $\pi ^{-} p$ .
In the spring of 1995 CHAOS, a new TRIUMF facility, began operating to
measure polarization $\pi^{\pm} p$ data below 100~MeV  ({\bf E560}),
and is expected to provide the
first such measurements below 70~MeV\cite{sm94p}.  A LAMPF experiment
({\bf E1178}) will measure analyzing powers between 45 and 265~MeV
for the charge exchange reaction in the fall of 1995\cite{co95p}.

The present solution (SM95) is compared with other recent VPI analyses
in Table~I. Here we display the quality of our fit to data in the
different charge channels, as well as the number of searched parameters
used in the fits.

\section{Formalism}
\label{sec:form}

\vskip .3cm
\centerline{\bf A. Chew-Mandelstam Formalism}
\vskip .1cm

Our energy-dependent partial-wave fits are parameterized in terms of
a coupled-channel Chew-Mandelstam K-matrix, as
described in Ref.\cite{ar85}. The elastic scattering amplitude for
each partial wave can be expressed in terms of a function $\bar K$
\begin{equation}
T_e = R_e \bar K / (1 - C_e \bar K)
\end{equation}
with
\begin{equation}
\bar K = K_e + C_i K^2_0 / (1 - C_i K_i) .
\end{equation}
Here $C_e$ and $C_i$ are the Chew-Mandelstam elastic ($\pi N$) and
inelastic ($\pi \Delta$) functions described in Ref.\cite{ar85};
the elastic phase space factor, $R_e$, is the imaginary part of $C_e$. In
order to control the behavior near the elastic threshold, the K-matrix
elements ($K_e$, $K_0$, and $K_i$) were expanded as polynomials in the
energy variable $Z = (W_{C.M.} - W_{th})$, where $W_{C.M.}$ and
$W_{th}$ are the center-of-mass and threshold energies, respectively,
for elastic pion-nucleon scattering.
Multiplying $K_0$ by an added factor of $Z$ allowed the fixing of
scattering lengths through the value of the leading term in $K_e$. It
should be noted that the above $\pi \Delta$ channel is a ``generic"
inelastic channel. As in previous analyses, the $S_{11}$ amplitude was
given an additional $\eta N$ coupling. Charge splitting was
accomplished through the multiplication of $\bar K$ by an appropriate
Coulomb barrier factor.

Single-energy analyses were parametrized as
\begin{equation}
S_e = (1 + 2iT_e) = \cos (\rho) \; e^{2 i \delta} ,
\end{equation}
with the phase parameters $\delta$ and $\rho$ expanded as
linear functions around the analysis energy, and with a slope (energy
derivative) fixed by the energy-dependent solution.

Details of the energy-dependent parameterization is as described
in Ref.\cite{ar85} with the following changes:
\begin{itemize}
\item
The subtraction point\cite{ar85}, $W_Z$,
is now $M+\mu - 500$~MeV ($M$ and $\mu$ being the nucleon and pion masses).
\item
All K-matrix elements were expanded as energy polynomials except for an
explicit K-matrix pole in the elastic component of the $P_{33}$ partial-wave.
\item
The $P_{33}$ was further modified for $\pi^- p$ and change-exchange by scaling
back the S-matrix modulus, $\eta$, to account for inverse pion-photoproduction
around the resonance. This is similar to the method used by Tromborg
et al.\cite{trom}.
\item
Once an appropriate hadronic amplitude was determined, charge corrections
were applied as described in Ref.\cite{ar85}.
\item
Threshold behavior was determined in the following manner.
The S-wave scattering lengths were linked to our dispersion relation
constraints, as described below. The P-wave scattering volumes were searched
without constraint. D-waves were softly constrained to the
Koch values\cite{koch},
and the higher waves were fixed to Koch's results\cite{koch}.
\end{itemize}

\vskip .3cm
\centerline{\bf B. Dispersion Relations Constraints}
\vskip .1cm

Constraints on the partial-wave fits were generated from the
forward $C^{\pm}$ amplitudes and the invariant $B$ amplitudes
at fixed-t in the range 0 to $-0.3$~(GeV/c)$^2$.
(As mentioned in Ref.\cite{ar94},
the $A^{\pm}$ dispersion relations, though not used as constraints, are
quite well satisfied.)
Reference \cite{ar94} describes our method of applying forward and
fixed-t dispersion relation constraints in order to generate solutions
with fixed values of the pion-nucleon coupling constant, $g^2/4\pi$,
and the isospin-even scattering length, $a^{(+)}$.
In the present work we have
generated a set of solutions in order to determine our sensitivity
to choices of the $\pi^- p$ scattering length and the pion-nucleon coupling
constant. Table~II displays the minimum  value of $\chi^2$ and
$g^2/4\pi$ found in fits with different choices for the $\pi^- p$ scattering
length, $a_{\pi^- p}$, and the integral
\begin{equation}
J_{GMO} = {1\over {4 \pi^2}} \int { {\sigma_{\pi^- p} - \sigma_{\pi^+ p} }
                             \over \omega} dk.
\end{equation}
which appears in the Goldberger-Miyazawa-Oehme (GMO) sum rule. Given a
value for the integral, $a^{(-)}$ is directly related to the chosen value
of $g^2/4\pi$.

Our final results were generated using $J_{GMO}$= $-1.05$~mb and
$a_{\pi^- p}$ = 0.085 $\mu^{-1}$. It is important to stress that
any reasonable set\cite{prelim} could be used and that the minimum value for
$g^2/4\pi$ depends only weakly upon the chosen values. Moreover, these
choices have a negligible effect on our results for the resonance
spectrum.

Given the above choices of $J_{GMO}$ and $a_{\pi^- p}$, Table~III shows
the sensitivity of our fits to the value of $g^2/4\pi$. The most
important difference between this mapping and our previous
result\cite{ar94} is the consistency of the optimal value of
$\pi NN$ coupling found from the constraints and all charge
channels. A problem once evident in Ref.\cite{ar94}, in the charge-exchange
channel, has now disappeared.

\vskip .3cm
\centerline{\bf C. Lesser Structure}
\vskip .1cm

There has been some criticism\cite{privcomm} of our method  of analysis,
based upon the absence
of some lesser (less than 4-star) structures in the VPI solutions.
It has been argued that this is the result of inflexibility in the
energy-dependent forms which we use. We have previously searched for
missing structure by iterating between single-energy and global fits,
examining each iteration for evidence of systematic deviations between the
resultant partial waves.

In order to explore this question more carefully, we have performed an
additional search
for (localized) missing structures, implementing the following strategy.
We have assumed a product S-matrix of the form
\begin{equation}
S = S_{FA93} \; S_P
\end{equation}
where $S_{FA93}$
is the solution\cite{ar94} used in our recent determinations of
$g^2/4\pi$, and $S_P$ was taken to have the form:
\begin{equation}
S_P = (1 + 2 i T_R) \left(  {{1+i K_B} \over {1-iK_B}} \right)
\end{equation}
with
\begin{equation}
T_R = { {\Gamma_{\pi N} /2} \over {W_R -W -i\Gamma/2} }
\end{equation}
where $\Gamma_{\pi N} = \rho_e \gamma_e$ and $\Gamma_i = \rho_i \gamma_i$.
The total width, $\Gamma$ is given by the sum of elastic ($\Gamma_{\pi N}$)
and inelastic ($\Gamma_i$) widths with phase-space factors, $\rho_{e,i}$,
normalized to unity at $W$ = $W_R$. In the above, $K_B$ is expressed as
$\gamma_B | T_R |^2$ (in order to keep the effect localized).

We mapped $\chi^2 (W_R , \gamma_B)$ for various combinations of
the constants $\gamma_e$ and $\gamma_i$. $W_R$ was varied from
1.4 to 2.3~GeV, in increments of 25~MeV, and $\gamma_B$ was varied from
$-10$ to $10$ in increments of 5. This was done for each partial wave.
A few candidates for extra structure were found in this way. Once
identified, these added structures were included in a fit constrained
by dispersion relations.

\vskip .3cm
\centerline{\bf D. Resonance Parameter Extraction}
\vskip .1cm

The resonance spectrum for our fit was extracted in the customary fashion.
A Breit-Wigner form plus background was used to fit partial-waves containing
structure over a selected range of energies. The precise form is given by
\begin{equation}
S=1+2iT=(1+2iT_R) \; \eta_B e^{2i\delta_B}
\end{equation}
with $T_R$ defined as above. The main requirement on the phase-space
factors is that $\rho_e$ should be proportional to $(W-M-\mu)^{l+1/2}$
at threshold, which allows for many possible choices.  For the background
we used
\begin{equation}
\delta_B = \delta_B^r + \alpha (W_R - W)
\end{equation}
with $\eta_B = \cos (\rho_B)$. To get initial values for the resonance
fitting, we implemented the speed plot (Speed = $|d T / d W |$)
advocated by H\"ohler\cite{ho92,ho94}. All 4-star
resonances show clear ``speed bumps" allowing the
extraction of initial parameters.

The values for extracted resonance parameters $(W_R, \Gamma_{\pi N}, \Gamma)$
were quite sensitive to the choice of phase-space factors, especially
for those resonances near threshold. For the $P_{33}$ in particular, it
was possible to obtain reasonable fits for a variety of assumed factors.
We ultimately adopted the form
\begin{equation}
\rho_e = \left( {q\over q_R} \right)^{2l+1}
 \left( {{q_R^2 + X^2}\over{q^2 + X^2}} \right)^l ,
\end{equation}
where $q$ and $q_R$ are the center-of-mass and resonance momenta.
This introduces a cutoff parameter, $X$, but seems to yield, for most 4-star
resonances, values consistent with previous
Particle Data Group (PDG)\cite{pdg} determinations.  We plan a more refined
analysis of the $P_{33}$(1232) resonance region once we receive the
data of Refs.\cite{pa93p}-\cite{co95p}. It is hoped that these new
measurements will help to resolve discrepancies existing is the current
database for this energy region.

\vskip .3cm
\centerline{\bf E. Complex Plane Mapping: Poles and Zeroes}
\vskip .1cm

Since the form used in our energy-dependent fits
can be analytically continued to
complex energies, it is straightforward to locate the complex energy
positions for the poles and zeroes which influence the on-shell behaviour
of the amplitudes.  We generate complex-plane contour plots of
$\ln ( |T|^2 )$ and pick a starting energy near the pole/zero. We then use
a Newton-Raphson algorithm to ``home in" on the structure. Results for
the pole positions (and residues) are given in the next section.

\section{Results of the Partial-Wave Analysis}
\label{sec:pwa}

The overall quality of our solution (SM95) is displayed in
Table~I, along with a number of our previous results.
Single-energy solutions were produced up to 2026~MeV. For these
single-energy solutions, starting values
for the partial-wave amplitudes and their (fixed) energy derivatives
were obtained from the energy-dependent fit. The scattering database
was supplemented with a constraint on each varied amplitude.
Constraint errors were taken to be 0.02 added in quadrature to 5\%
of the amplitude. Such constraints are essential to prevent the solutions
from `running away' when a bin is sparsely populated with scattering
data, but have little effect when sufficient data exists.
In Table~IV we compare the energy-dependent and single-energy fits
to the data. These solutions are displayed graphically in Fig.~2. Here we
also compare with the previous solution SM90. Some of the largest
changes are seen in $S_{11}$ (near the $\eta$-cusp), in $P_{13}$ (at
intermediate energies), and in $P_{11}$ (at higher energies).

Our search for lesser structures, as described in Section III.C,
revealed only three possibilities for obtaining a significantly
improved fit. After inclusion into the main analysis,
we determined that only two
of these lesser structures, in the $S_{11}$ and $F_{15}$ partial waves,
remained significant enough to keep in our final fit.
These can be seen as small
``bumps" on the high-energy shoulders of the $S_{11}(1650)$ and
$F_{15}(1680)$ resonances. The $S_{11}$ structure is also evident in
the speed plot of Fig.~3.

Pole positions and the associated
Breit-Wigner parameters are presented graphically in Fig.~4, and are
listed in Tables~V and VI. PDG values are also given for comparison.
We have not attempted to associate the added
structures in $S_{11}$ and $F_{15}$ with any specific PDG designation.
A structure found in $P_{13}$ was likewise left ``unnamed".

We are able to resolve all 4-star structures listed by the
PDG within our energy range. We also determined structure
in the speed plots for $P_{33}$ around 1800~MeV, and
for $P_{31}$  near 1400~MeV. Neither of these were
resolvable via a Breit-Wigner fit.
The difficulty with these unresolved structures can be seen in Fig.~5,
which reveals a rather complicated interference
between nearby zeroes and poles. Many of the weaker structures appear
as pole-zero combinations, with a zero lying between the pole and
the physical axis.

\section{Comparisons and Discussion}
\label{sec:rs}

As we find structures associated with all 4-star resonances in our energy
range, we can claim qualitative agreement with the Karlsruhe, CMB,
and Kent State analyses. The $P_{13}$ result is difficult to interpret. We
find a pole position close to the CMB value but the Breit-Wigner fit results
in a resonance energy between the 4-star $P_{13}(1720)$ and 1-star
$P_{13}(1910)$.

Our two additional resonances, found in sweeping a Breit-Wigner
form through the partial-wave amplitudes, could possibly be related to PDG
1- and 2-star resonances found previously in the $S_{11}$ and $F_{15}$
amplitudes. The Karlsruhe group reported a structure (denoted as the
$F_{15}(2000)$ 2-star resonance) at 1882~MeV, not  far from our value.
The elasticity we found is also similar to that found by the Karlsruhe
and Kent State
groups. The next $S_{11}$ resonance reported above the $S_{11}(1650)$
is the 1-star $S_{11}(2090)$. Our structure appears about 150~MeV
below this. It is interesting to note that H\"ohler\cite{ho94} found a
similar structure in his speed plot of the KA84 solution.

The PDG 3-star $D_{13}(1700)$ resonance is not evident in the present analysis.
The Kent State group found an elasticity consistent
with zero for this resonance.
The photo-couplings to the $D_{13}(1700)$  are also consistent with zero in
the most recent PDG estimates. If this resonance exists, it remains very
difficult to detect. We do see the 3-star $P_{33}(1600)$, though our pole
position is quite different from the Karlsruhe and CMB values. The resonance
energy estimates, from the Karlsruhe, CMB and
Kent State groups, also span a wide range.

In summary, we have found that our present analysis gives all the dominant
structures found in earlier works, along with a couple of new ones which
may be related to previous 1- or 2-star states. We also found
the value of $g^2/4\pi$
to be more consistently determined by individual charge channels
and the constraints than was the case in our first
set\cite{ar94} of $\chi^2$ maps. These amplitudes will be used
as input for our upcoming analysis of pion photoproduction data.
Results for the new $S_{11}$ and $F_{15}$ resonances
will be especially interesting, as these states
presently have no assigned photo-coupling estimates
in the Review of Particle Properties.

This reaction is incorporated into the SAID program\cite{tel}, which is
maintained at Virginia Tech.

\acknowledgments

The authors express their gratitude to I.G. Alekseev, A. Badertscher,
B. Bassalleck, J.T. Brack, D.V. Bugg, J.-P. Egger,
E. Friedman, G. Jones, V.V. Kulikov,
I.V. Lopatin, C.A. Meyer, R. Minehart, B.M.K. Nefkens, D. Po\v{c}ani\'c,
R.A. Ristinen, M. Sadler, M. Sevior, G.R. Smith, and J. Stasko
for providing experimental data prior to publication or
clarification of information already published.
We also acknowledge useful communications with G. H\"ohler.
I.S. acknowledges the hospitality extended by the Physics Department of
Virginia Tech.
This work was supported in part by the U.S.~Department of Energy Grant
DE--FG05--88ER40454 and a NATO Collaborative Research Grant 921155U.


\newpage
{\Large\bf Figure captions}\\
\newcounter{fig}
\begin{list}{Figure \arabic{fig}.}
{\usecounter{fig}\setlength{\rightmargin}{\leftmargin}}
\item
{Energy-angle distribution of recent (post-1990) (a) $\pi ^{-}p$, (b)
$\pi ^{+}p$, and (c) charge exchange data.
$\pi ^{-}p$ data are [observable (number of data)]:
d$\sigma$/d$\Omega$~(291), P~(308), partial total cross sections (21),
R~(45), and A~(45).
$\pi ^{+}p$ data are:
d$\sigma$/d$\Omega$~(169), $\sigma ^{t}$~(51), P~(56), R~(41), and A~(41).
Charge exchange data are: d$\sigma$/d$\Omega$~(24), total cross sections
$\sigma ^{tot}$~(6), P~(23).  Total cross sections are plotted at zero
degrees.}
\item
{Partial-wave amplitudes (L$_{2I, 2J}$) from 0 to 2.1~GeV.  Solid (dashed)
curves give the real (imaginary) parts of amplitudes corresponding to the
SM95 solution.  The real (imaginary) parts of single-energy solutions are
plotted as filled (open) circles. The previous SM90 solution \cite{ar91} is
plotted with long dash-dotted (real part) and short dash-dotted (imaginary
part) lines.  The dotted curve gives the value of Im~T - T$^*$~T.  All
amplitudes have been multiplied by a factor of 10$^3$ and are dimensionless.}
\item
{Speed plot of the $S_{11}$ amplitude. The solid (dashed) line gives the
result for solution SM95 (SM90)~[1].}
\item
{Comparison of complex plane and Breit-Wigner fits for resonances found in
solution SM95. Complex plane poles are plotted as stars (the boxed star
denotes a second-sheet pole). W$_{\rm R}$ and W$_{\rm I}$ give real and
imaginary parts of the center-of-mass energy. The total (elastic) widths
are denoted by narrow (wide) bars for each resonance. (a) S- and P-wave
resonances, (b) D- and F-wave resonances, (c) G- and H-wave resonances.}
\item
{Complex plane pole/zero plot for the (a) $S_{11}$, (b) low-energy $P_{11}$,
(c) high-energy $P_{11}$, and
(d) $P_{33}$ partial-wave amplitudes. P and Z denote the pole and zero
positions. S indicates a second-sheet pole.
Stars locate nearby PDG resonance positions and the underlying bars
give the PDG values for the elastic and full widths.}
\end{list}

\newpage
\mediumtext
\vfill
\eject
Table~I. Comparison of present (SM95) and previous (FA93, SM90, and
FA84) energy-dependent partial-wave analyses of elastic $\pi^{\pm} p$
scattering and charge-exchange (CXS) data.  $N_{prm}$ is the number
parameters (I = 1/2 and 3/2) varied in the fit.
\vskip 10pt
\centerline{
\vbox{\offinterlineskip
\hrule
\hrule
\halign{\hfill#\hfill&\qquad\hfill#\hfill&\qquad\hfill#\hfill
&\qquad\hfill#\hfill&\qquad\hfill#\hfill&\qquad\hfill#\hfill
&\qquad\hfill#\hfill\cr
\noalign{\vskip 6pt} %
Solution&Range~(MeV)&$\chi^2$/$\pi ^{+}p$~data&$\chi^2$/$\pi ^{-}p$~data&
$\chi^2$/CXS~data&$N_{prm}$&Ref. \cr
\noalign{\vskip 6pt}
\noalign{\hrule}
\noalign{\vskip 10pt}
SM95 & $0 -2100$ & 22593/10197 & 18855/9421 & 4442/1625 &94/80& Present \cr
\noalign{\vskip 6pt}
FA93 & $0 -2100$ & 23552/10106 & 20747/9304 & 4834/1668 &83/77& \cite{ar94} \cr
\noalign{\vskip 6pt}
SM90 & $0 -2100$ & 24897/10031 & 24293/9344 &10814/2132 &76/68& \cite{ar91} \cr
\noalign{\vskip 6pt}
FA84 & $0 -1100$ &  7416/ 3771 & 10658/4942 & 2062/ 717 &64/57& \cite{ar85} \cr
\noalign{\vskip 10pt}}
\hrule}}
\vfill
\eject
Table~II. Table of $\chi^2$ values for different minimal values of
$g^2/4\pi$. The value of $\chi^2_{min}$ is given for different values of
the GMO integral ($J_{GMO}$) and the $a_{\pi^- p}$ scattering length.
\vskip 10pt
\centerline{
\vbox{\offinterlineskip
\hrule
\hrule
\halign{\hfill#\hfill&\qquad\hfill#\hfill&\qquad\hfill#\hfill
&\qquad\hfill#\hfill\cr
\noalign{\vskip 6pt}
  $g^2/4\pi_{min}$ &  $3 \; a_{\pi^- p}$  &   $J_{GMO}$ &  $\chi^2_{min}$\cr
\hbox{\hskip .1cm} & ($\mu^{-1}$) & ($mb$) & \hbox{\hskip .1cm}\cr
\noalign{\vskip 6pt}
\noalign{\hrule}
\noalign{\vskip 6pt}
 13.71 &  0.250 &  $-1.00$  &   46,381 \cr
\noalign{\vskip 10pt}
 13.75 &  0.250 &  $-1.05$  &   46,241 \cr
\noalign{\vskip 10pt}
 13.78 &  0.250 &  $-1.10$  &   46,370 \cr
\noalign{\vskip 10pt}
\noalign{\hrule}
\noalign{\vskip 6pt}
 13.72 &  0.255 &  $-1.00$  &   46,340 \cr
\noalign{\vskip 10pt}
 13.76 &  0.255 &  $-1.05$  &   46,236 \cr
\noalign{\vskip 10pt}
 13.81 &  0.255 &  $-1.10$  &   46,386 \cr
\noalign{\vskip 10pt}
\noalign{\hrule}
\noalign{\vskip 6pt}
 13.73 &  0.260 &  $-1.00$  &   46,287 \cr
\noalign{\vskip 10pt}
 13.77 &  0.260 &  $-1.05$  &   46,221 \cr
\noalign{\vskip 10pt}
 13.81 &  0.260 &  $-1.10$  &   46,422 \cr
\noalign{\vskip 10pt}}
\hrule}}
\vfill
\eject
Table~III. Table of $\chi^2$ values for different choices of the
pion-nucleon coupling used in the analysis of $\pi^{\pm} p$ elastic
scattering and charge-exchange (CXS) data. The number of data (or
constraints) is given in brackets.
\vskip 10pt
\centerline{
\vbox{\offinterlineskip
\hrule
\hrule
\halign{\hfill#\hfill&\qquad\hfill#\hfill&\qquad\hfill#\hfill
&\qquad\hfill#\hfill&\qquad\hfill#\hfill&\qquad\hfill#\hfill
&\qquad\hfill#\hfill\cr
\noalign{\vskip 6pt}
\noalign{\hrule}
\noalign{\vskip 6pt}
 Solution & $g^2/4\pi$ &  Data  &   Constraints & $\pi^+$ &  $\pi^-$  & CXS\cr
\hbox{\hskip .1cm} & \hbox{\hskip .1cm} & (21220) & (496) & (10190) & (9350)
& (1680) \cr
\noalign{\vskip 6pt}
\noalign{\hrule}
\noalign{\vskip 6pt}
 E337 & 13.37 &   47921 &  709 &  23269 &  19935 &   4717\cr
\noalign{\vskip 6pt}
 E350 & 13.50 &   46776 &  527 &  22759 &  19466 &   4551\cr
\noalign{\vskip 6pt}
 E363 & 13.63 &   46127 &  410 &  22557 &  19108 &   4462\cr
\noalign{\vskip 6pt}
 E375 & 13.75 &   45919 &  352 &  22599 &  18877 &   4443\cr
\noalign{\vskip 6pt}
 E387 & 13.87 &   46030 &  367 &  22799 &  18775 &   4456\cr
\noalign{\vskip 6pt}
 E400 & 14.00 &   46483 &  452 &  23176 &  18789 &   4518\cr
\noalign{\vskip 6pt}
\noalign{\hrule}
\noalign{\vskip 6pt}
$\chi^2_{min}$ & \hbox{\hskip .1cm} & 45918  &  355 & 22552 & 18766
&  4435\cr
\noalign{\vskip 6pt}
$g^2/4\pi_{min}$ & \hbox{\hskip .1cm} & 13.77 & 13.79 & 13.69 & 13.93
& 13.77\cr
\noalign{\vskip 6pt}
$\Delta (g^2/4\pi_{min})$ & \hbox{\hskip .1cm} & 0.01 &  0.03 &  0.02
&  0.02 &   0.03\cr
\noalign{\vskip 10pt}}
\hrule}}
\vfill
\eject
Table~IV. Single-energy (binned) fits of combined $\pi ^{\pm}p$ elasic
scattering and charge-exchange data, and $\chi^2$ values.  $N_{prm}$ is the
number parameters varied in the single-energy fits, and $\chi^2_E$ is given
by the energy-dependent fit, SM95, over the same energy interval.
\vskip 10pt
\centerline{
\vbox{\offinterlineskip
\hrule
\hrule
\halign{\hfill#\hfill&\qquad\hfill#\hfill&\qquad\hfill#\hfill
&\qquad\hfill#\hfill&\qquad\hfill#\hfill&\qquad\hfill#\hfill
&\qquad\hfill#\hfill\cr
\noalign{\vskip 6pt} %
T$_{lab}$~(MeV)&Range~(MeV)&$N_{prm}$&$\chi^2$/data&$\chi^2_E$&&\cr
\noalign{\vskip 6pt}
\noalign{\hrule}
\noalign{\vskip 10pt}
  30 & $  26 -  33 $ &  4 & 242/136 &  290 &&\cr
\noalign{\vskip 6pt}
  47 & $  45 -  49 $ &  4 &  72/81  &  108 &&\cr
\noalign{\vskip 6pt}
  66 & $  61 -  69 $ &  4 & 189/122 &  245 &&\cr
\noalign{\vskip 6pt}
  91 & $  89 -  92 $ &  4 &  79/73  &   98 &&\cr
\noalign{\vskip 6pt}
 124 & $ 121 - 126 $ &  6 &  74/61  &   88 &&\cr
\noalign{\vskip 6pt}
 145 & $ 141 - 147 $ &  6 &  36/42  &   50 &&\cr
\noalign{\vskip 6pt}
 170 & $ 165 - 174 $ &  6 &  87/67  &   95 &&\cr
\noalign{\vskip 6pt}
 193 & $ 191 - 194 $ &  6 &  45/54  &   52 &&\cr
\noalign{\vskip 6pt}
 217 & $ 214 - 220 $ &  6 &  69/59  &  152 &&\cr
\noalign{\vskip 6pt}
 238 & $ 235 - 240 $ &  6 &  79/72  &   95 &&\cr
\noalign{\vskip 6pt}
 266 & $ 262 - 270 $ &  7 & 117/88  &  163 &&\cr
\noalign{\vskip 6pt}
 292 & $ 291 - 293 $ &  8 & 148/129 &  222 &&\cr
\noalign{\vskip 6pt}
 309 & $ 306 - 310 $ &  8 & 169/140 &  227 &&\cr
\noalign{\vskip 6pt}
 334 & $ 332 - 335 $ &  9 &  96/58  &  133 &&\cr
\noalign{\vskip 6pt}
 352 & $ 351 - 352 $ &  9 &  79/110 &  148 &&\cr
\noalign{\vskip 6pt}
 389 & $ 387 - 390 $ &  9 &  30/28  &  101 &&\cr
\noalign{\vskip 6pt}
 425 & $ 424 - 425 $ & 10 & 146/139 &  206 &&\cr
\noalign{\vskip 6pt}
 465 & $ 462 - 467 $ & 15 & 355/120 &  466 &&\cr
\noalign{\vskip 6pt}
 500 & $ 499 - 501 $ & 15 & 159/136 &  185 &&\cr
\noalign{\vskip 6pt}
 518 & $ 515 - 520 $ & 16 & 101/79  &  149 &&\cr
\noalign{\vskip 6pt}
 534 & $ 531 - 535 $ & 19 & 134/128 &  203 &&\cr
\noalign{\vskip 6pt}
 560 & $ 557 - 561 $ & 19 & 331/151 &  570 &&\cr
\noalign{\vskip 6pt}
 580 & $ 572 - 590 $ & 19 & 369/286 &  460 &&\cr
\noalign{\vskip 6pt}
 599 & $ 597 - 600 $ & 22 & 250/151 &  502 &&\cr
\noalign{\vskip 6pt}
 625 & $ 622 - 628 $ & 23 & 126/95  &  199 &&\cr
\noalign{\vskip 6pt}
 662 & $ 648 - 675 $ & 23 & 584/352 &  750 &&\cr
\noalign{\vskip 6pt}
 721 & $ 717 - 725 $ & 25 & 203/169 &  300 &&\cr
\noalign{\vskip 6pt}
 745 & $ 743 - 746 $ & 25 & 164/100 &  293 &&\cr
\noalign{\vskip 6pt}
 765 & $ 762 - 767 $ & 26 & 190/169 &  330 &&\cr
\noalign{\vskip 6pt}
 776 & $ 774 - 778 $ & 26 & 226/155 &  318 &&\cr
\noalign{\vskip 10pt}}
\hrule}}
\vfill
\eject
Table~IV (continued).
\vskip 10pt
\centerline{
\vbox{\offinterlineskip
\hrule
\hrule
\halign{\hfill#\hfill&\qquad\hfill#\hfill&\qquad\hfill#\hfill
&\qquad\hfill#\hfill&\qquad\hfill#\hfill&\qquad\hfill#\hfill
&\qquad\hfill#\hfill\cr
\noalign{\vskip 6pt} %
T$_{lab}$~(MeV)&Range~(MeV)&$N_{prm}$&$\chi^2$/$\pi N$~data&$\chi^2_E$&&\cr
\noalign{\vskip 6pt}
\noalign{\hrule}
\noalign{\vskip 10pt}
 795 & $ 793 - 796 $ & 26 & 206/165 &  319 &&\cr
\noalign{\vskip 6pt}
 820 & $ 813 - 827 $ & 26 & 398/304 &  482 &&\cr
\noalign{\vskip 6pt}
 868 & $ 864 - 870 $ & 32 & 277/195 &  407 &&\cr
\noalign{\vskip 6pt}
 888 & $ 886 - 890 $ & 33 & 173/144 &  309 &&\cr
\noalign{\vskip 6pt}
 902 & $ 899 - 905 $ & 34 & 550/416 &  852 &&\cr
\noalign{\vskip 6pt}
 927 & $ 923 - 930 $ & 36 & 240/200 &  373 &&\cr
\noalign{\vskip 6pt}
 962 & $ 953 - 971 $ & 36 & 384/299 &  557 &&\cr
\noalign{\vskip 6pt}
1000 & $ 989 -1015 $ & 38 & 689/423 &  865 &&\cr
\noalign{\vskip 6pt}
1030 & $1022 -1039 $ & 39 & 284/272 &  400 &&\cr
\noalign{\vskip 6pt}
1044 & $1039 -1049 $ & 40 & 357/243 &  538 &&\cr
\noalign{\vskip 6pt}
1076 & $1074 -1078 $ & 43 & 221/218 &  427 &&\cr
\noalign{\vskip 6pt}
1102 & $1099 -1103 $ & 44 & 226/173 &  335 &&\cr
\noalign{\vskip 6pt}
1149 & $1147 -1150 $ & 44 & 325/210 &  459 &&\cr
\noalign{\vskip 6pt}
1178 & $1165 -1192 $ & 44 & 763/394 &  985 &&\cr
\noalign{\vskip 6pt}
1210 & $1203 -1216 $ & 44 & 286/233 &  372 &&\cr
\noalign{\vskip 6pt}
1243 & $1237 -1248 $ & 44 & 452/283 &  641 &&\cr
\noalign{\vskip 6pt}
1321 & $1304 -1337 $ & 44 & 728/401 &  950 &&\cr
\noalign{\vskip 6pt}
1373 & $1371 -1375 $ & 44 & 308/166 &  581 &&\cr
\noalign{\vskip 6pt}
1403 & $1389 -1417 $ & 44 & 547/408 &  783 &&\cr
\noalign{\vskip 6pt}
1458 & $1455 -1460 $ & 44 & 280/258 &  448 &&\cr
\noalign{\vskip 6pt}
1476 & $1466 -1486 $ & 44 & 486/323 &  648 &&\cr
\noalign{\vskip 6pt}
1570 & $1554 -1586 $ & 46 & 831/546 & 1125 &&\cr
\noalign{\vskip 6pt}
1591 & $1575 -1606 $ & 46 & 425/336 &  647 &&\cr
\noalign{\vskip 6pt}
1660 & $1645 -1674 $ & 46 & 553/391 &  821 &&\cr
\noalign{\vskip 6pt}
1720 & $1705 -1734 $ & 46 & 398/279 &  528 &&\cr
\noalign{\vskip 6pt}
1753 & $1739 -1766 $ & 46 & 660/439 &  863 &&\cr
\noalign{\vskip 6pt}
1838 & $1829 -1845 $ & 46 & 461/290 &  709 &&\cr
\noalign{\vskip 6pt}
1875 & $1852 -1897 $ & 46 & 989/682 & 1358 &&\cr
\noalign{\vskip 6pt}
1929 & $1914 -1942 $ & 46 & 840/501 & 1297 &&\cr
\noalign{\vskip 6pt}
1970 & $1962 -1978 $ & 46 & 477/271 &  688 &&\cr
\noalign{\vskip 6pt}
2026 & $2014 -2037 $ & 46 & 414/320 &  794 &&\cr
\noalign{\vskip 10pt}}
\hrule}}
\vfill
\eject
Table~V. Masses, half-widths ($\Gamma$/2), and values for
($\Gamma _{\pi N}$/$\Gamma$) are listed for isospin 1/2 baryon resonances,
along with associated pole positions from our solution SM95~(second sheet
poles are denoted by a $\dagger$). Corresponding residues are given as a
modulus and phase (in degrees).  Average values from the Review of Particle
Properties\cite{pdg} are given in square brackets.
\vskip 10pt
\centerline{
\vbox{\offinterlineskip
\hrule
\hrule
\halign{\hfill#\hfill&\qquad\hfill#\hfill&\qquad\hfill#\hfill
&\qquad\hfill#\hfill&\qquad\hfill#\hfill&\qquad\hfill#\hfill
&\qquad\hfill#\hfill\cr
\noalign{\vskip 6pt} %
Resonance &$W_{R}$&$\Gamma$/2&$\Gamma _{\pi N}$/$\Gamma$&Pole &Residue&\cr
\noalign{\vskip 6pt}
(* rating)& (MeV)&   (MeV)  &                     &(MeV)&(MeV, $^{\circ}$)&\cr
\noalign{\vskip 6pt}
\noalign{\hrule}
\noalign{\vskip 10pt}
P$_{11}$(1440)& 1467 &   220    &     0.68           & $1346-i88$ &
(42, -101)&\cr
\noalign{\vskip 6pt}
   & & & & ($1383-i105$)$\dagger$ & (92, -54)$\dagger$&\cr
\noalign{\vskip 6pt}
****          &[1440]&  [175]   &        [0.65]      & & &\cr
\noalign{\vskip 10pt}
D$_{13}$(1520)& 1515 &    53    &         0.61       & $1515-i55$ &
(34, 7) &\cr
\noalign{\vskip 6pt}
****          &[1520]&   [60]   &        [0.55]      & & &\cr
\noalign{\vskip 10pt}
S$_{11}$(1535)& 1535 &    33    &         0.31       & $1501-i62$ &
(31, -12)&\cr
\noalign{\vskip 6pt}
****          &[1535]&   [75]   &        [0.45]      & & &\cr
\noalign{\vskip 10pt}
S$_{11}$(1650)& 1667 &    45    & $\approx$1.0       & $1673-i41$ &
(22, 29)&\cr
\noalign{\vskip 6pt}
****          &[1650]&   [75]   &        [0.70]      & & &\cr
\noalign{\vskip 10pt}
S$_{11}$ & 1712      &    92    &         0.27       & $1689-i96$ &
(72, -85)&\cr
\noalign{\vskip 10pt}
D$_{15}$(1675)& 1673 &    77    &         0.38       & $1663-i76$ &
(29, -6) &\cr
\noalign{\vskip 6pt}
****          &[1675]&   [75]   &        [0.45]      & & &\cr
\noalign{\vskip 10pt}
F$_{15}$(1680)& 1678 &    63    &         0.68       & $1670-i60$ &
(40, 1)&\cr
\noalign{\vskip 6pt}
****          &[1680]&   [65]   &        [0.65]      & & &\cr
\noalign{\vskip 10pt}
P$_{11}$(1710)&  --- &   ---    &     ---            & $1770-i189$ &
(37, -167)&\cr
\noalign{\vskip 6pt}
****          &[1710]&  [50]   &        [0.15]       & & &\cr
\noalign{\vskip 10pt}
P$_{13}$ & 1820 &   177        &         0.16        & $1717-i194$&
(39,-70)&\cr
\noalign{\vskip 10pt}
F$_{15}$ & 1814 &    88        &         0.10        & $1793-i94$ &
(27, -56)&\cr
\noalign{\vskip 10pt}}
\hrule}}
\vfill
\eject
\vfill
\eject
Table~V (continued).
\vskip 10pt
\centerline{
\vbox{\offinterlineskip
\hrule
\hrule
\halign{\hfill#\hfill&\qquad\hfill#\hfill&\qquad\hfill#\hfill
&\qquad\hfill#\hfill&\qquad\hfill#\hfill&\qquad\hfill#\hfill
&\qquad\hfill#\hfill\cr
\noalign{\vskip 6pt} %
Resonance &$W_{R}$&$\Gamma$/2&$\Gamma _{\pi N}$/$\Gamma$&Pole &Residue&\cr
\noalign{\vskip 6pt}
(* rating)& (MeV)&   (MeV)  &                     &(MeV)&(MeV, $^{\circ}$)&\cr
\noalign{\vskip 6pt}
\noalign{\hrule}
\noalign{\vskip 10pt}
G$_{17}$(2190)& 2131 &   238    &         0.23       & $2030-i230$&
(46, -23)&\cr
\noalign{\vskip 6pt}
****          &[2190]&  [225]   &        [0.15]      & & &\cr
\noalign{\vskip 10pt}
H$_{19}$(2220)& 2258 &   167    &         0.26       & $2203-i268$&
(68, -43)&\cr
\noalign{\vskip 6pt}
****          &[2220]&  [200]   &        [0.15]      & & &\cr
\noalign{\vskip 10pt}
G$_{19}$(2250)& 2291 &   386    &         0.10       & $2087-i340$&
(24, -44)&\cr
\noalign{\vskip 6pt}
****          &[2250]&  [200]   &        [0.10]      & & &\cr
\noalign{\vskip 10pt}}
\hrule}}
\vfill
\eject
\vfill
\eject
Table~VI. Parameters for isospin 3/2 baryon resonances.  Notation
as in Table~V.
\vskip 10pt
\centerline{
\vbox{\offinterlineskip
\hrule
\hrule
\halign{\hfill#\hfill&\qquad\hfill#\hfill&\qquad\hfill#\hfill
&\qquad\hfill#\hfill&\qquad\hfill#\hfill&\qquad\hfill#\hfill
&\qquad\hfill#\hfill\cr
\noalign{\vskip 6pt} %
Resonance &$W_{R}$&$\Gamma$/2&$\Gamma _{\pi N}$/$\Gamma$&Pole &Residue&\cr
\noalign{\vskip 6pt}
(* rating)& (MeV)&   (MeV)  &                     &(MeV)&(MeV, $^{\circ}$)&\cr
\noalign{\vskip 6pt}
\noalign{\hrule}
\noalign{\vskip 10pt}
P$_{33}$(1232)& 1233 &    57    &     $\approx$1.0   & $1211-i50$ &
(38, -22)&\cr
\noalign{\vskip 6pt}
****          &[1232]&   [60]   &       [0.994]      & & &\cr
\noalign{\vskip 10pt}
P$_{33}$(1600)&  --- &    ---   &         ---        & $1675-i193$&
(52, 14)&\cr
\noalign{\vskip 6pt}
***           &[1600]&  [175]   &       [0.17]       & & &\cr
\noalign{\vskip 10pt}
S$_{31}$(1620)&  1617 &  54    &         0.29        & $1585-i52$&
(14, -121)&\cr
\noalign{\vskip 6pt}
****          &[1620]&   [75]   &       [0.25]       & & &\cr
\noalign{\vskip 10pt}
D$_{33}$(1700)& 1680 &   136    &        0.16        & $1655-i121$&
(16, -12)&\cr
\noalign{\vskip 6pt}
****          &[1700]&   [150]  &       [0.15]       & & &\cr
\noalign{\vskip 10pt}
F$_{35}$(1905)& 1850 &    147   &        0.12        & $1832-i127$&
(12, -4)&\cr
\noalign{\vskip 6pt}
****          &[1905]&   [175]  &       [0.10]       & & &\cr
\noalign{\vskip 10pt}
P$_{31}$(1910)& 2152  &   380   &        0.26        & $1810-i247$&
(53, -176)&\cr
\noalign{\vskip 6pt}
****          &[1910]&   [125]  &       [0.22]       & & &\cr
\noalign{\vskip 10pt}
D$_{35}$(1930)& 2056 &    295   &        0.11        & $1913-i123$ &
(8, -47)&\cr
\noalign{\vskip 6pt}
***           &[1930]&   [175]  &       [0.15]       & & &\cr
\noalign{\vskip 10pt}
F$_{37}$(1950)& 1921 &    116   &        0.49        & $1880-i118$&
(54, -17)&\cr
\noalign{\vskip 6pt}
****          &[1950]&   [150]  &       [0.38]       & & &\cr
\noalign{\vskip 10pt}}
\hrule}}

\vfill
\eject
\end{document}